\documentclass[epj]{svjour}

\usepackage[dvips]{epsfig}
\usepackage{hyperref}
\usepackage{color}

\usepackage{txfonts}

\newcommand{\rem}[1]{}

\begin{document}

\title{Atomistic simulations of the sliding friction of graphene flakes}

\author{Federico Bonelli\inst{1} \and
Nicola Manini\inst{1} \and
Emiliano Cadelano\inst{2} \and
Luciano Colombo\inst{2}}
\institute{%
Physics Department and INFM - University of Milan
and European Theoretical Spectroscopy Facility\\
Via Celoria 16, 20133 Milano, Italy
\and
Department of Physics, University of Cagliari and
SLACS-INFM/CNR Sardinian Laboratory for Computational Materials Science\\
Cittadella Universitaria, I-09042 Monserrato (Ca), Italy
}

\date{April 2, 2008}

\abstract{
Using a tight-binding atomistic simulation, we simulate the recent
atomic-force microscopy experiments probing the slipperiness of graphene
flakes made slide against a graphite surface.
Compared to previous theoretical models, where the flake was assumed to be
geometrically perfect and rigid, while the substrate is represented by a
static periodic potential, our fully-atomistic model includes quantum
mechanics with the chemistry of bond breaking and bond formation, and the
flexibility of the flake.
These realistic features, include in particular the crucial role of the
flake rotation in determining the static friction, in qualitative agreement
with experimental observations.
}


\maketitle

\section{Introduction}
\label{intro}

The scanning tunneling microscope (STM) \cite{Binning82}, and even more the
atomic-force microscope (AFM) \cite{Binning86}, have triggered perhaps the
biggest wave of advances and discoveries ever in surface science and
nanoscience.
Experimental investigations of friction on the atomic scale have become
possible by virtue of the friction force microscope (FFM).
In a FFM experiment a sharp tip scans a sample surface with atomic
precision, while lateral forces are recorded with a resolution that can
reach the pN range.
Since Mate {\it et al.}\ \cite{Mate87} investigated the nanoscale periodic
frictional force map of a graphite surface using a tungsten tip, many
studies have been conduced experimentally and theoretically.
In recent works the Leiden group \cite{Dienwiebel04,Dienwiebel05} has
probed quantitatively the well known slipperiness of graphite, responsible
for its excellent lubrication properties.
Morita {\it et al.}\ \cite{Morita96} suggested that in FFM experiments on
layered materials, such as $\rm MoS_2$ or graphite, a flake, consisting of
several hundred atoms in contact with the substrate, can attach to the tip.
By controlling the relative angles of individual nanoflakes to achieve a
suitable lattice mismatch, thus incommensurate contact \cite{Hirano90}, an
almost frictionless sliding was demonstrated for dry and wearless tip-surface
contact, a phenomenon known as superlubricity.
Several experiments and calculations an have been probing the effects of
lattice mismatch on friction
\cite{Maier07,Maier08,Filleter09,Vanossi06,Vanossi07PRL,Castelli08Lyon},
showing that incommensuracy often prevents global ingraining of large
areas, thus attenuating the consequent strongly dissipative stick-slip
motion.
Theoretically, atomic-scale friction on ideal solid surfaces is often
described by simple balls-and-springs models such as the Tomlinson model
\cite{Tomlinson29}, where a single atom, or a more structured tip
\cite{Tomanek91}, is dragged through a spring along a static periodic
potential energy surface.
In the present work we implement a Tomlinson-like model of a simulated FFM
experiment where dissipation of a finite graphene flake is driven along a
graphite substrate, in a fully atomistic scheme based on a tight-binding
(TB) force field.
Compared to similar models in the literature
\cite{Sorensen96,Verhoeven04,Fusco04,Filippov08}, where the flake is
assumed to be perfect and rigid and the substrate is represented by an
analytically defined static periodic potential, our fully-atomistic TB
simulation explores two realistic features, namely:
(i) The flake-substrate interaction potential is not classical and contains
quantum mechanics with the chemistry of bond breaking and bond formation.
(ii) The flake is nonrigid, so that during its advancement it can  deform
and relax.
In Sec.~\ref{model} we introduce the model implementation details.
Section~\ref{results} reports the results obtained, in particular for the
friction dependency on the flake size, the rotation angle relative to the
substrate, and the applied load; we compare the results of the present
model to experiment and to previous calculations.
The final Sec.~\ref{conclusion} discusses the results and the advantages
and drawbacks of the present model.

\section{\label{model} The model}

We describe the sliding by means of a generalized Tomlinson-like model
similar to that of Ref.~\cite{Verhoeven04}, but including the following
features:
(i) the interaction among all carbon atoms is realized in terms of 
the tight-binding scheme of Xu {\it et al.}\ \cite{Xu92}, and 
(ii) the flake is not rigid but is allowed to deform and rotate while sliding.
Interatomic forces are computed as customary in the TB scheme
\cite{Colombo05}; the hopping parameters and the pairwise repulsive
potential term follow the scaling form given by Xu {\it et al.}\ \cite{Xu92}.
All interatomic interactions vanish at a cutoff distance $r_c=2.6$~\AA.
This distance sits in between the nearest-neighbor and the
next-nearest-neighbor distances of carbon atoms of the equilibrium sp$^3$
diamond structure, and of the sp$^2$ graphene plane.
It is also shorter than the interlayer distance of graphite, which is as
long as 3.35~\AA\ \cite{Zacharia04}.
The present TB model has been applied successfully to investigate several
low-dimensional carbon systems \cite{Canning97,Yamaguchi07,Cadelano09}.
In particular, this parameterization reproduces the experimental
equilibrium distance $d_{\rm graph} = 1.4224$~\AA\ of the graphitic plane.
To study friction, we use a model constituted by a graphene flake sliding
over a single infinite rigid graphene sheet.
The infiniteness of the substrate is simulated by repeating periodically a
regular arrangement of $N^{\rm sub}$ atoms at positions $\vec r^{\rm
  \,sub}_i$ in the $x-y$ plane.
We consider a periodic rectangular supercell, containing as many carbon
atoms as necessary for the flake not to interact with its periodic images.
\begin{figure} 
\centering
\epsfig{file=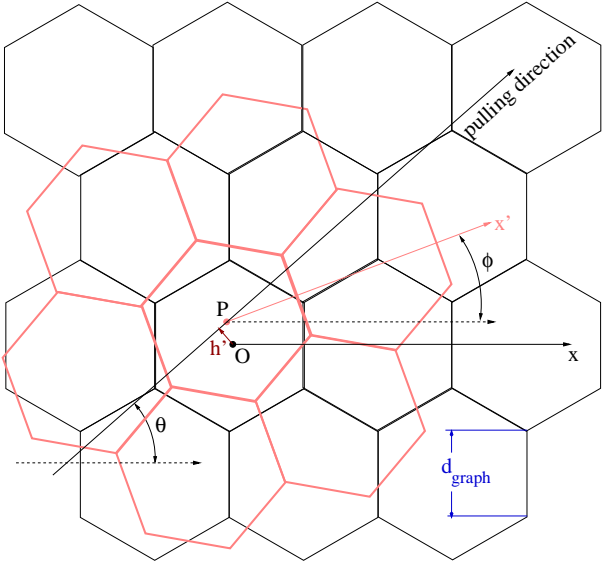,width=80mm,clip=}
\caption{(Color online) \label{thetaphi}
The definition of the angle $\phi$ measuring the initial rotation of the
flake (thick, pink lines) with respect to the substrate honeycomb structure
(thin, black lines).
The pulling line is defined by its distance $h'$ to the center of a
substrate hexagon and the pulling angle $\theta$.
To take as a reference the  AB stacking ($\phi=0$), we define $h'$ in terms of
the parameter $h=h' + \frac 12 d_{\rm graph}$.
}
\end{figure}

We have studied three different regular hexagonal flakes composed of
$N^{\rm fl} = 24$, 54, and 96 atoms, respectively.
Initially, the flake is rotated by an angle $\phi$ with respect to the
substrate and translated horizontally to put its center of mass along a
sliding line at a distance $h'$ from the center of a substrate hexagon.
As illustrated in Fig.~\ref{thetaphi}, this line is oriented at an angle
$\theta$ from the $\hat x$ direction, which is defined by being
parallel to the zig-zag direction of the honeycomb lattice.
In our simulations we drag the flake along this pulling line and usually
let the flake atoms relax in all directions ($x$, $y$ and $z$), whereas the
substrate remains completely rigid.
The $\phi$ angle defines the stacking mismatch, which has a central
importance for friction.
Specific values of $\theta$ such as $0^\circ$ and $30^\circ$ define special
pulling directions where the pulled flake encounters periodic corrugations,
while aperiodic corrugations are experienced for generic $\theta$ angles.
The range of interest of both angles $\theta$ and $\phi$ runs from
$0^\circ$ to $30^\circ$: outside this range we recover equivalent
geometries.

\begin{figure} 
\centering
\epsfig{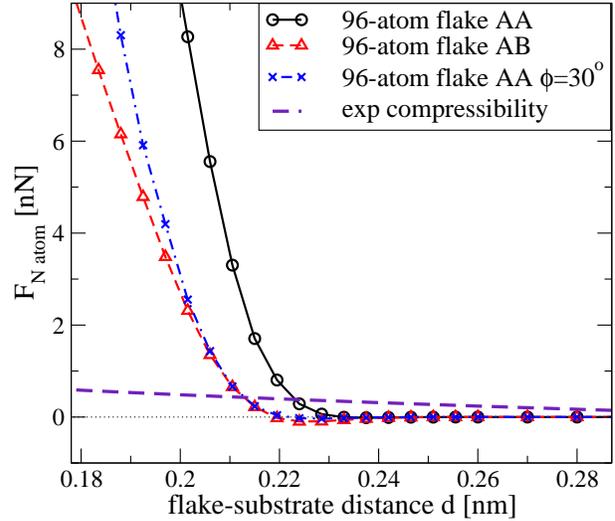}
\caption{(Color online) \label{Fnflakerigid}
The computed normal force per atom $F_{N\;\rm atom}$ as a function of the
fixed rigid flake-substrate distance $d$.
The curves refer to different stackings (AA or eclipsed, AB and an
incommensurate obtained starting from AB rotated by $\phi = 30^\circ$)
for a 96-atom flake undeformed flake.
The compression curve based on the low-pressure data of
Ref.~\cite{Yeoman69} is reported for comparison.
}
\end{figure}

To simulate an AFM experiment we introduce a constant load $F_N$ pushing
the flake against the substrate along the vertical $z$ direction and
simulating the force applied by the actual tip.
This load acts against the reaction forces produced by the interaction
with the substrate.
These forces are reported in Fig.~\ref{Fnflakerigid} for a $N^{\rm
  fl}=96$-atom flake in several configurations, and of course vanish beyond
$r_c=2.6$~\AA, the TB interaction cutoff.
For a distance $d\geq 0.18$~nm, the load force per atom does not exceed
10~nN (a total load in the tens to few hundred nN for a flake composed by
10 to $10^2$ atoms, corresponding to a load pressure $\simeq 4$~Mbar),
which is a value practically accessible to FFM experiments
\cite{Maier07,Maier08,Filleter09}.
A load per atom of 0.5~nN, withing the selected force-field model, produces
an approach distance near 0.21~nm, similar to the one obtained by assuming
for the flake-substrate system the equilibrium interlayer separation of
graphite (3.35~\AA) and the very soft $c$-axis compressibility
\cite{Yeoman69}, expressed as $d \ln c/dP = -2.8 \times
10^{-6}$~bar$^{-1}$.
Note however that we apply this compressibility relation, also sketched in
Fig.~\ref{Fnflakerigid} for distances and pressures that go beyond its
linear-response range of validity: in the region of close approach, $d\leq
0.22$~nm, the actual force is likely to increase more rapidly, like in the
TB model curves.
We implement a Tomlinson-like dynamics with each flake atom coupled
horizontally to a rigid ``support'' by elastic springs.
The support is a set of ideal graphene-net points, which coincide with the
initial flake atomic positions: $\vec{r}_i^{\rm \,sup}=\vec{r}_i^{\rm
 \, fl}(t=0)$.
This support is then translated rigidly parallel to the substrate.
Its orientation is fixed once and for all by the angle $\phi$.
The support advances by steps of length $\delta x$ along the direction
defined by the pulling angle $\theta$ and lateral shift $h'$.
After a few tests, we select an advancement step $\delta x = 0.0024$~nm.
After each advancement step relaxation, we evaluate and store the total
spring energy and the total dragging force, defined as follows:
\begin{eqnarray}\label{en_spring}
E^{\rm spr}&=&
\frac K2 \sum_{i=1}^{N^{\rm fl}}
\left(r^{\rm fl}_{x\,i}- r^{\rm sup}_{x\,i}\right)^2 +
\left(r^{\rm fl}_{y\,i}- r^{\rm sup}_{y\,i}\right)^2
\,;
\\\label{f_spring}
\vec {F}_{\varparallel}^{{\rm \,spr}}&=&
\sum_{i=1}^{N^{\rm fl}} \vec {F}_{\varparallel\,i}^{{\rm \,spr}}
\,,
\\\label{f_spring2}
\vec {F}_{\varparallel\,i}^{{\rm \,spr}}&=&
- K \left(\vec {r}_i^{\rm \,fl}-\vec r^{\rm \,sup}_i\right)_\varparallel
\,,
\end{eqnarray}
where the $\varparallel$ symbol indicates the in-plane component.
The component $F^{{\rm \,spr}}_{\varparallel\,i}$ of $\vec
{F}_{\varparallel\,i}^{{\rm \,spr}}$ along the pulling direction equals the
force needed to make the support advance, so that the work of this force
equals $F^{{\rm \,spr}}_{\varparallel\,i}\, \delta x$, for an infinitesimal
displacement in the pulling direction.

In an AFM experiment, the scanning tip speed is typically of the order of tens
or hundreds nm/s, much slower than the fast dynamics of the flake.
Assuming that the substrate temperature is fairly low, it is appropriate to
consider a quasi-static motion of the flake as follows:
after each advancement of the support, all flake atomic positions are made
relax in all directions by damped dynamics \footnote{
The advantage of damped dynamics with respect to a standard
energy-minimization algorithm is that at each step it relaxes smoothly and
predictably to the nearest local-minimum configuration.
On the other hand, this algorithm is computationally less efficient than,
e.g., a conjugated-gradient technique.
Eventually we settle on a fairly fast converging damped dynamics with a
time step $\delta t=4$~fs.
}
under the combined action of (i) the TB forces, (ii) the vertical load
force $F_N$, and (iii) the horizontal spring forces that attract the flake
atoms near the support points.
When the stationary equilibrium position is reached (defined by no force
component exceeding a threshold of $10^{-2}$~nN), the support moves one
step further and the whole relaxation procedure is repeated until the
support reaches the end of a pre-defined path.
With the selected fairly small advancement step $\delta x$, each relaxation
requires typically 10 to 200 MD steps.
We consider a path of moderate length $d\simeq 1$~nm divided into
approximately 400 advancement steps.
The spring elastic constant coupling the support and the flake in the
$x$-$y$ plane is a quite critical parameter of the present model.
Softer springs allow the flake a greater freedom to translate, rotate, and
deform, with better pinning to energetically favorable sites and more
pronounced stick-slip motion and higher friction.
Harder springs enforce a more stiff flake showing little or no stick-slip
motion.
The limit $K\to\infty$ matches the model by Verhoeven {\it et
  al.}\ \cite{Verhoeven04}.
The spring constant mimic the combined interaction between the flake and
the tip and the elastic tip response.
We suggest that a value
\begin{equation}
\label{Kiniz}
K
=
0.5\,\frac{\rm eV}{\AA^2}
\simeq
8.01 \, \frac{\rm N}{\rm m}
\end{equation}
corresponding to about 10\% of the stretching stiffness of a carbon-carbon
bond within a graphene layer should probably be fairly realistic.
We also perform computations with softer ($K=0.1 $~eV/\AA$^{2}$) and harder
($K_x=K_y=2.5$~eV/\AA$^{2}$) springs, as detailed in Sec.~\ref{results}.

\begin{figure} 
\centering
\epsfig{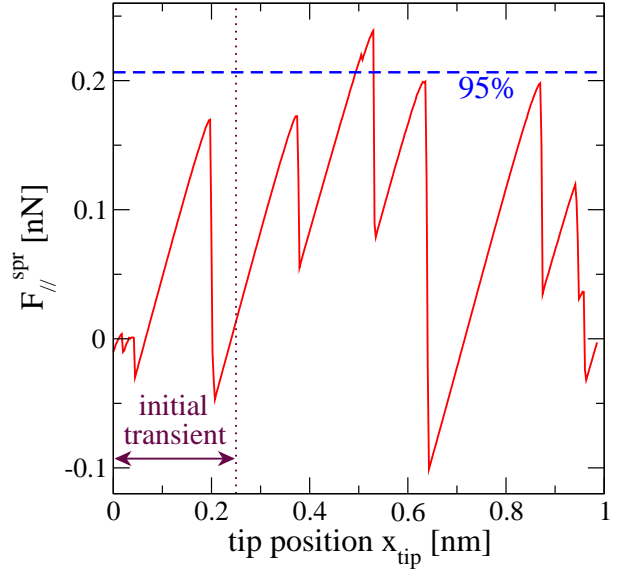}
\caption{(Color online) \label{Fstaticth}
The spring total lateral force component $\vec {F}_{\varparallel}^{{\rm
    \,spr}}$ projected along the scan line, for a 24-atom flake dragged
starting from an initial stacking AB along a pulling angle $\theta =
15^\circ$, and with a stacking angle $\phi=30^\circ$.
The spring constant is $K= 0.1$~eV/\AA$^{2}$, the total load $F_N=100$~nN.
The $95 \%$ force level (dashed line) estimates the static friction force
$F_{\rm fric}$.
}
\end{figure}
In our calculations, by convention, we estimate the static friction force
$F_{\rm fric}$ along a given sliding path by the force level below which
95\% of the spring force values $F_{\varparallel}^{\rm spr}$ encountered
along the path (at the end of each relaxation, and excluding an initial
transient) are found.
This definition is illustrated in Fig.~\ref{Fstaticth}.
\begin{figure} 
\centering
\epsfig{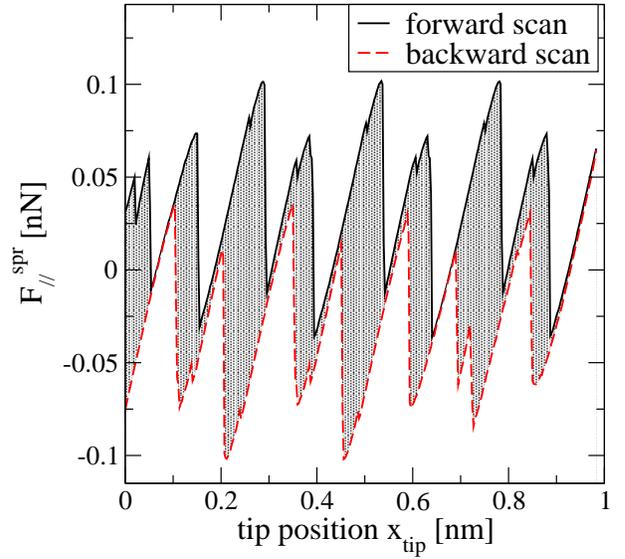}
\caption{\label{A+R} (Color online)
The dynamic-friction force is evaluated as the energy dissipated through a
forward-backward scan (solid and dashed lines respectively).
The shaded area between the two curves measures the energy dissipated by
friction.
The conditions are as follows: $N^{\rm fl}=24$, $F_N=100$~nN,
$\theta=0^\circ$, $\phi=0^\circ$, $K = 0.1$~eV/\AA$^{2}$.
}
\end{figure}
The dynamic-friction force is defined in terms of the energy dissipated in
a forward-backward loop.
Figure~\ref{A+R} illustrates this concept: the component of the spring
force parallel to the advancement direction shows a clear hysteretic
behavior in a forward-backward scan.
The shaded area enclosed between the two curves measures the
energy $E_{\rm fric}$ dissipated by friction, and is clearly related to the
stick-slip events.
The average dynamical friction force $F_{\rm fric}^{\rm dynamic}= E_{\rm
  fric}/d \simeq 0.325$~nN is of course smaller than the static friction
force $F_{\rm fric}\simeq0.9$~nN evaluated according to the 95\% protocol
described above.
In the following we will focus on the static friction force $F_{\rm fric}$,
which is cheaper to compute and eventually of the same order as
$F_{\rm fric}^{\rm dynamic}$.
As the flake advances along its path, it deforms and rotates around its
center of mass.
In particular, to understand the evolution of the static friction force, it
is useful to track the flake instantaneous stacking angle $\phi_A$, which
generally differs from the fixed support angle $\phi$.
At each relaxed configuration, we calculate $\phi_A$ as an average over all
flake atoms $i$, of a sine projection obtained as the length of the vector
product of two vectors in the horizontal $xy$ plane: ${\bf R}_i^{\rm cm}$,
joining the flake center of mass to the $i$-th flake atom, and the
corresponding vector ${\bf R}_i^{0 \, \rm cm}$ computed for the unrotated
$\phi=0^{\circ}$ support.
Explicitly:
\begin{equation}
\label{phiact}
\phi_A=
\arcsin\!\left(
\sum_i \frac{R_i^{{\rm cm} \ y} R_i^{0 \, {\rm cm} \ x}-R_i^{{\rm cm} \ x}
    R_i^{0 \, {\rm cm} \ y}}
{N^{\rm fl} \, |{\bf R}_i^{{\rm cm}}| \, |{\bf R}_i^{0 \, {\rm cm}}|}
\right).
\end{equation}

\section{\label{results} Results}

We analyze the friction force dependence on several physical parameters,
namely the pulling angle $\theta$, stacking angle $\phi$, applied load,
flake size and position of the scan-line.
Experimentally, the number of flake atoms, estimated in the order of 100,
is not well determined, while the total applied load and the total force
acting on the flake are under control in the FFM.
For ease of comparison with experiments, our discussion shall always deal
with total quantities, i.e.\ summed over the flake atoms.

\subsection{Relaxation to equilibrium}

\begin{figure} 
\centering
\epsfig{file=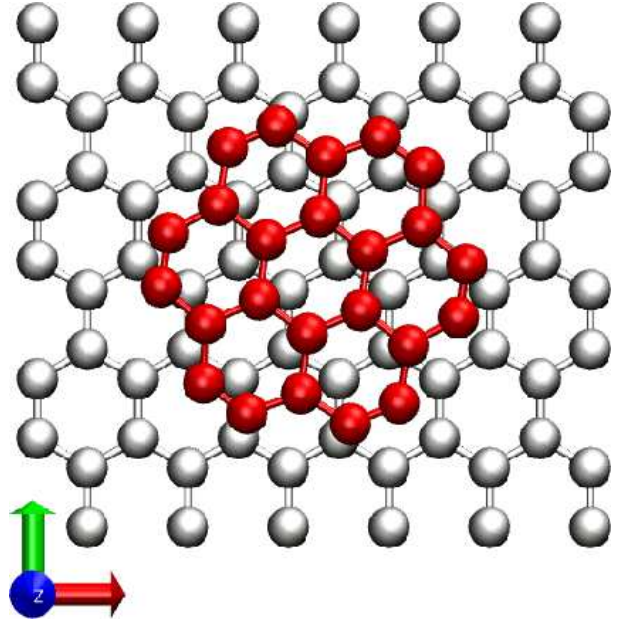,width=80mm,clip=}
\caption{(Color online) \label{relax24}
The relaxed configuration of the $24$-atom flake. Static substrate atoms
are clear (white), flake atoms are dark (red).
} 
\end{figure}

\begin{figure} 
\centering
\epsfig{file=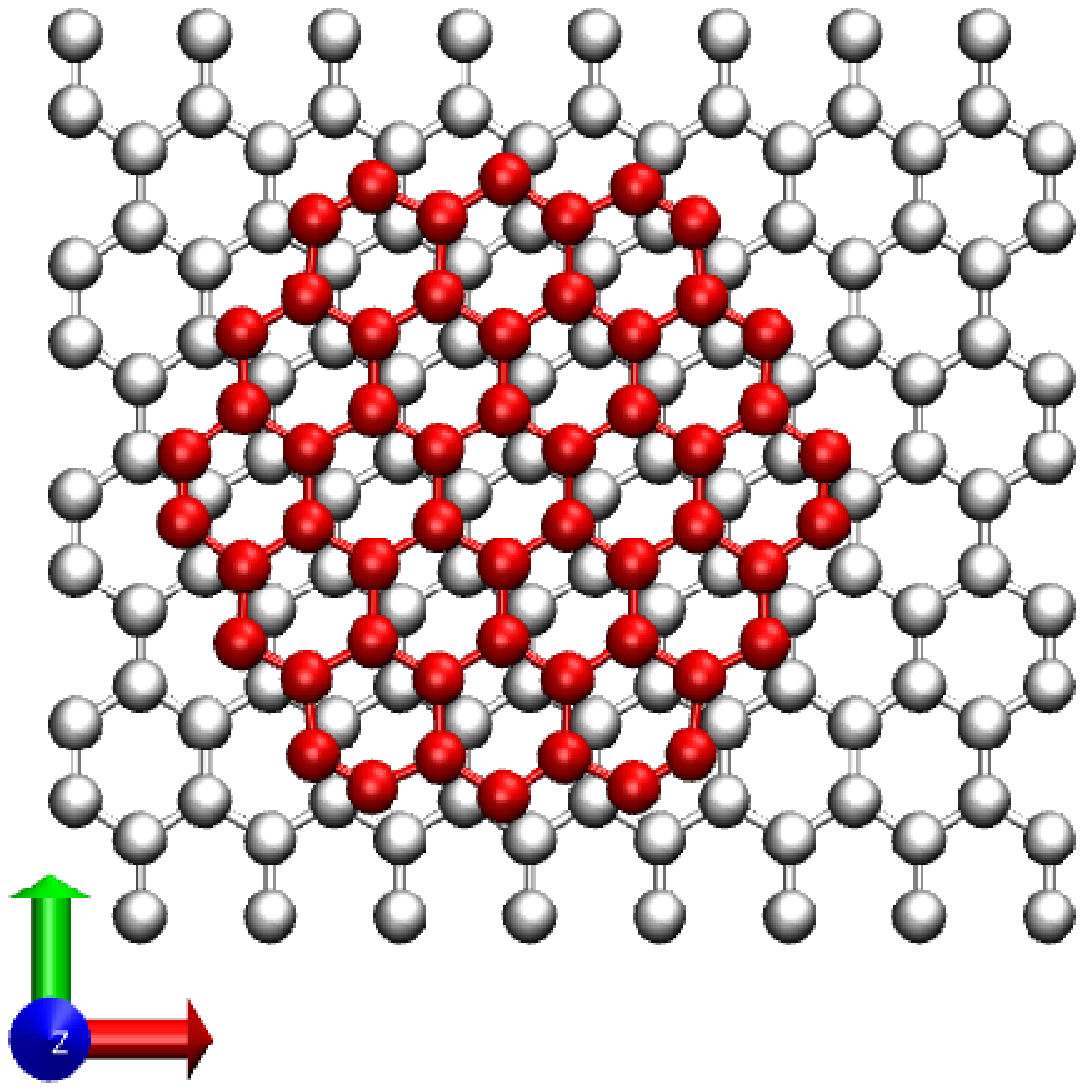,width=80mm,clip=}
\caption{(Color online) \label{relax54}
The relaxed configuration of the $54$-atom flake. Static substrate atoms
are clear (white), flake atoms are dark (red).
}
\end{figure}

\begin{figure} 
\centering
\epsfig{file=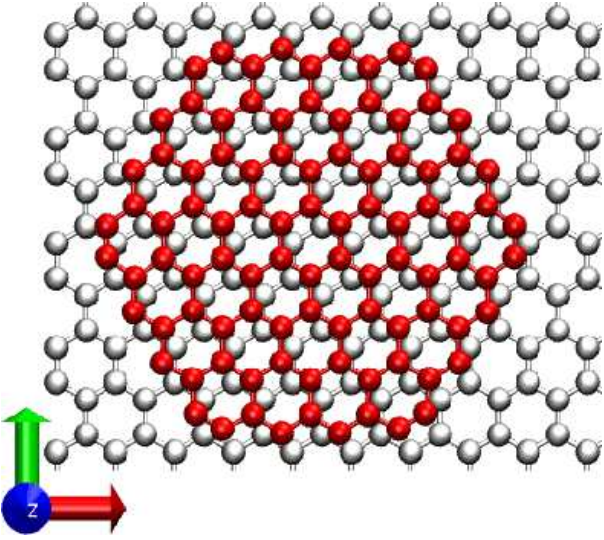,width=80mm,clip=}
\caption{(Color online) \label{relax96}
The relaxed configuration of the $96$-atom flake. Static substrate atoms
are clear (white), flake atoms are dark (red).
}
\end{figure}

Optimally stacked configurations are important in providing the most
efficient sticking points during a friction sliding experiment.
Figs. \ref{relax24}, \ref{relax54}, and \ref{relax96} report typical such
relaxed configurations for $24-$, $54-$ and $96-$atom flakes respectively,
obtained under the action of a total load of $100$ nN, and with no spring
connection to a support ($K=0$).
The relaxed configuration (up to a symmetry rotation/translation) depends
only moderately on the starting stacking, unless the initial stacking angle
is strongly incommensurate.
The average vertical flake-substrate separation is 0.192~nm.
The equilibrium configuration tends to arrange the flake so as to minimize
the number of flake atoms stacked on top of a substrate atom: indeed
Fig.~\ref{Fnflakerigid} shows that the eclipsed ``AA'' stacking produces
the strongest repulsive force at the same distance.
Geometrically different non-optimal configurations are characterized by
typical excess total energies of 1~eV or less.

\subsection{The stick-slip movement}

\begin{figure} 
\centering
\epsfig{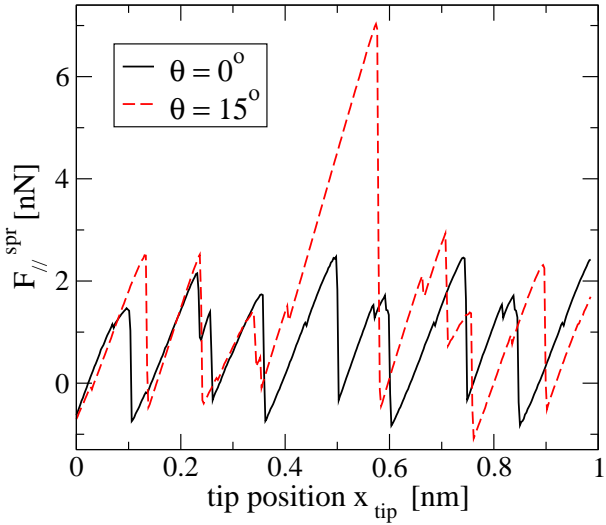}
\caption{(Color online) \label{FlatVSxtipth0-15}
The component of the springs force $F_{\varparallel}^{\rm spr}$ in the
dragging direction as a function of the support advancement distance
$x_{\rm tip}$ for two different pulling directions: $\theta=0^\circ$
(solid) and $\theta=15^\circ$ (dashed).
The simulation involves a $24$ atom flake with total applied load of
100~nN, support stacking angle $\phi=0^\circ$ and spring constant
$K=0.1$~eV/\AA$^2$.
}
\end{figure}

We come now to the actual simulation of sliding friction: to start with,
Fig.~\ref{FlatVSxtipth0-15} displays the pulling force
$F_{\varparallel}^{\rm spr}$ measured along two sliding paths of different
commensurability nature: $\theta=0^\circ$ (periodic) and $\theta=15^\circ$
(aperiodic).
As expected, the regular stick-slip pattern of the $\theta=0^\circ$ path is
replaced by an irregular dependency in the $\theta=15^\circ$ trajectory.
The initial part of the $\theta=0^\circ$ trajectory is not periodic because
of the usual startup transient behavior: the first 0.2~nm are omitted from
the calculation of the friction force, as discussed above.

\begin{figure} 
\centering
\epsfig{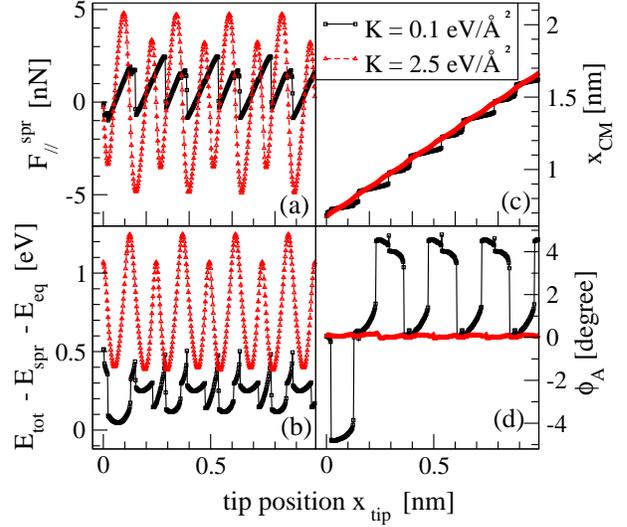}
\caption{(Color online) \label{AllVSxtip}
Several physical quantities plotted as functions of the support position
$x_{\rm tip}$: 
(a) the parallel component of the springs lateral force
$F_{\varparallel}^{\rm spr}$, (b) the flake excess energy $E^{\rm
  tot}-E^{\rm spr}-E^{\rm eq}$, (c) the flake center-of-mass advancement
$x_{\rm CM}$ along the pulling direction, and (d) the flake instantaneous
rotation angle $\phi_A$ relative to the substrate.
Two different values of the spring constant are compared: soft,
$K=0.1$~eV/\AA$^2$ (solid, squares), and hard, $K=2.5$~eV/\AA$^2$ (dashed,
triangles).
These simulations involve $N^{\rm fl}=24$, $F_N=100$~nN,
$\theta=0^\circ$, and $\phi=0^\circ$.
}
\end{figure}

The stick-slip motion and the role of the spring stiffness is understood
even better by comparing other physical quantities with
$F_{\varparallel}^{\rm spr}$.
In particular, Fig.~\ref{AllVSxtip} displays the internal energy of the
flake-substrate interaction, the displacement of the flake center of mass
along the pulling direction, and the actual stacking angle $\phi_A$.
We focus initially on the solid curves, obtained in a simulation based on
soft springs with $K=0.1$~eV/\AA$^2$.
After the initial transient, where the flake explores once a configuration
with a negative $\phi_A$, it then jumps back and forth between two kinds of
sticking configurations: the most favorable one characterized by
$\phi_A\simeq 4^\circ$, and another one, at 0.1 to 0.2~eV higher energy,
near $\phi_A\simeq 0^\circ$.
These stick-slip jumps overcome energy barriers whose heights are of order
0.2 to 0.3~eV.
This energy amplitude sets the temperature range of validity of the present
zero-temperature calculations to a few hundred Kelvin: when
thermally-activated slips through energy barriers do not have enough time
to occur, i.e.\ for a not-too-small tip advancement speed
\cite{Riedo03,Gnecco03}, these slips are unlikely and our estimates of
friction should be fairly reliable.
Stick-slip events occur with correlated jumps in the spring force, flake
excess energy, flake position, and stacking angle.
The very different dashed curves show that stiff springs (with
$K=2.5$~eV/\AA$^2$) produce a much stronger and more rigid binding of the
flake to the rigid support.
Accordingly, such an unrealistically rigid coupling suppresses the stick-slip
behavior: both the advancement, shown in Fig.~\ref{AllVSxtip}(c), and the
spring force, Fig.~\ref{AllVSxtip}(a), become smooth and jump-less.
Despite the suppression of stick-slip, we observe higher force peaks for
the stiffer springs, thus indicating a higher static friction than for the
softer springs.
This is due to the flake being forced to cross high potential-energy
barriers, Fig.~\ref{AllVSxtip}(b), with little or no possibility to avoid
them by (i) shifting away from the pulling direction, (ii) deforming, and
(iii) ``rotating around'' ($|\phi_A|<0.2^\circ$).
A similar behavior is observed also in different geometries.
The spring strength tuning the coupling between the flake and the AFM tip
plays therefore an important role in the model calculations.
We checked that up to spring constants of an intermediate value
($K=0.5$~eV/\AA$^2$) our model still performs a stick-slip motion similar
to the one found in experiment \cite{Dienwiebel04} (where the cantilever
harmonic constants values was estimated $K\simeq 0.36$~eV\,/\AA$^{2}$),
especially in the fully commensurate $\phi=0^\circ$ stacking.

\subsection{Flake-size effects}

\begin{figure} 
\centering
\epsfig{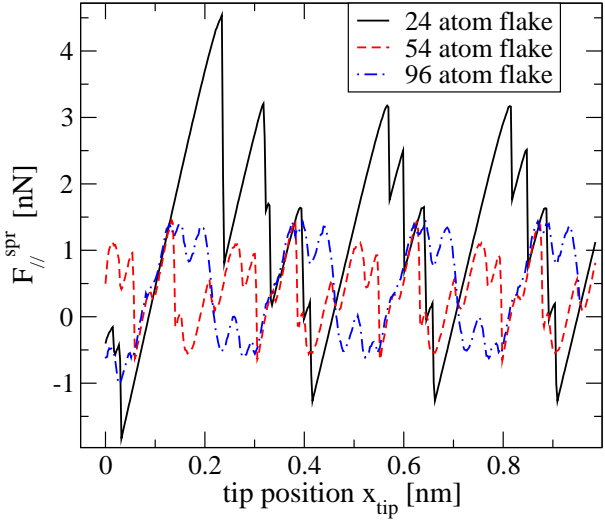}
\caption{(Color online) \label{24-54-96}
The parallel component of springs lateral force, $F_{\varparallel}^{\rm
  spr}$, as a function of the support position $x_{\rm tip}$ for three
flake sizes: 24-atom flake (solid), 54-atom flake atoms (dashed), and
96-atom flake atoms (dot-dashed).
The three simulations involve a total applied load $F_N=100$~nN,
$\theta=0^\circ$, $\phi=30^\circ$, and soft springs constants
$K=0.1$~eV/\AA$^2$.
}
\end{figure} 

Figure \ref{24-54-96} compares the frictional behavior of flakes of
different size, showing that friction tends to decrease with increasing
flake size.
In detail, the static friction force of the 3 flakes is $F_{\rm
  fric}=2.99$~nN, 1.38~nN, and 1.24~nN for the 24, 54, and 96-atom flakes
respectively.
This decrease is not surprising, since the more reactive atoms at the flake
boundary tend to bend down toward the substrate.
As a result, friction is dominated by these boundary atoms, which amount to
75\% of the 24-atom flake but only 44\% of the 96-atom flake.
Moreover, the 96-atom flake advances continuously, and shows no
stick-slip, at variance with the 24-atom and 54-atom flakes.
This difference is due to a reduced rotational freedom of the 96-atom flake
flake, due to coupling to the support acting at a larger distance from the
flake center.
The flake rotational freedom, i.e.\ the angular range of $\phi_A$ explored
around $\phi$, does represent a key issue in the friction physics of carbon
flakes sliding over a graphite surface, as pointed out by Filippov {\it et
  al.} \cite{Filippov08}.

In the framework of our model, with each atom tied to the moving support by
an individual spring, the flake can both shift normally to the pulling
direction and rotate around its center of mass: these degrees of freedom
(and, more weakly, the possibility of the flake to distort) affect friction
in two very different manners depending on the contact being commensurate
or incommensurate.
When the flake is pulled at a commensurate stacking,
e.g.\ $\phi=0^{\circ}$, it encounters high potential-energy barriers thus
high friction: the combined possibility of rotations and lateral shifts
allows the flake dribble the high barriers through local changes of
trajectory.
The first effect of flake shifts and rotations is then to reduce the
friction of highly commensurate contacts. 
In contrast, when the flake slides with an incommensurate stacking,
e.g.\ $\phi=15^{\circ}$, it does not encounter high-energy barriers nor
efficiently binding configurations, thus producing a low-friction motion.
However rotations and shifts allow the flake to locate deeper energy wells
(both moving apart from the pulling direction and rotating so as to explore
different stacking configurations), where the flake can stick, eventually
providing sizeable friction.
This second effect is therefore to raise the friction for incommensurate
contacts and eventually destroy superlubricity, as was observed and
discussed by Filippov {\it et al.}\ \cite{Filippov08}.
In our simple model the single parameter $K$ tunes the flake rotational
freedom and that of shifting perpendicular to the pulling direction.
However, while its effect on the translational freedom is independent of
size, the rotational freedom does depend on the larger torque that springs
of the same stiffness exert on flake atoms more remote from the flake
center, as is also to be expected for a flake sticking to a not too sharp
AFM tip.

\begin{figure} 
\centering
\epsfig{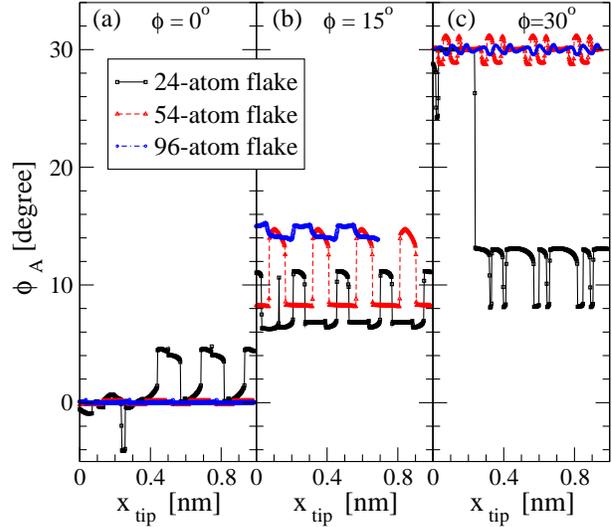}
\caption{(Color online) \label{angVSxtip24-54}
The instantaneous rotation angle $\phi_A$ as a function of the support
position $x_{\rm tip}$ for three values of the support stacking angle: (a)
$\phi=0^{\circ}$, (b) $\phi=15^{\circ}$, and (c) $\phi=30^{\circ}$.
Three flake sizes are considered: $N^{\rm fl}=24$ (black solid line),
$N^{\rm fl}=54$ (red dashed line) and $N^{\rm fl}=96$ (blue
dot-dashed line).
Simulations are carried out with total applied load of $100$~nN, pulling
angle $\theta=0^{\circ}$ and soft springs $K=0.1$~eV/\AA$^2$.
}
\end{figure}

This point shows clearly in Fig.~\ref{angVSxtip24-54}, which displays the
evolution of the actual rotation angle $\phi_A$ along the scanline, for
three different support stacking angles and three different flake sizes:
$N^{\rm fl}=24$, $54$, and $96$.
Rotational fluctuations decrease as the flake size increases.
Indeed significant systematic deviations of $\phi_A$ from $\phi$ are
apparent in many cases, especially $N^{\rm fl}=24$.
In particular, the small 24-atom flake for $\phi=30^\circ$ rotates all the
way to $\phi_A\leq 10^\circ$, thus displaying angular oscillations in
excess of $15^{\circ}$, for an average angle $\langle \phi_A
\rangle\simeq16^\circ$, very different from $\phi$.
When plotting the dependence of friction force $F_{\rm fric}$ on the
stacking angle, it will make more sense to use, instead of the initial
stacking angle $\phi$, the average flake rotation angle $\langle \phi_A
\rangle$, although even this indicator does not account properly for
rotational fluctuations.
This large rotational freedom, for hard springs, is almost completely
frozen:
in that case the largest rotational fluctuation we observe is as little as
$\simeq 3^{\circ}$ for $N^{\rm fl}=24$ and $\phi=15^\circ$.

\begin{figure} 
\centering
\epsfig{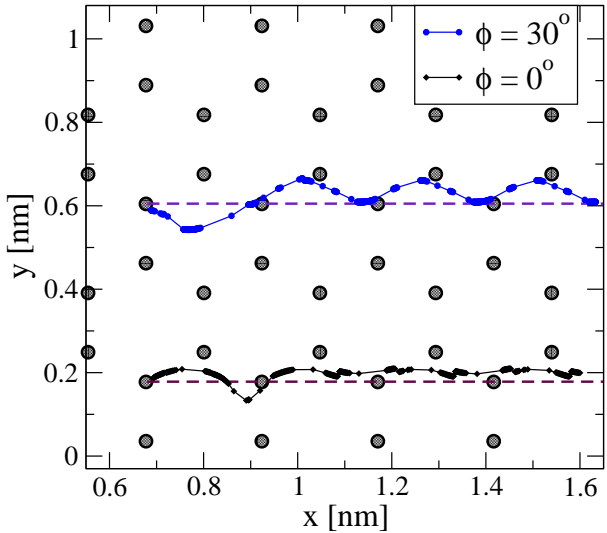}
\caption{(Color online) \label{trajectory}
Subsequent points marking the trajectories of the center of mass of a
24-atom flake in the $x$-$y$ plane for $\phi=30^{\circ}$ (incommensurate
stacking, blue circles) and for $\phi=0^{\circ}$ (commensurate stacking,
black diamonds) at the end of each relaxation cycle.
The dashed lines represent the support scanlines ($\theta=0^{\circ}$);
large circles represent the substrate atomic positions.
The simulations are the soft-spring ones of Fig.~\ref{AllVSxtip}.
}
\end{figure}

As for lateral shifts, for soft springs $K=0.1$~eV/\AA$^2$ we observe
shifts perpendicular to the pulling line of the order of 1~\AA, depicted in
Fig.~\ref{trajectory} which displays the actual path followed by the center
of mass of a 24-atom flake pulled by the support along a $\theta=0^{\circ}$
scanline.
Note that, similarly to rotational fluctuations, the incommensurate
stacking $\phi=30^{\circ}$ yields larger lateral shifts than the
commensurate stacking $\phi=0^{\circ}$.
With hard springs the possibility of the flake to perform lateral shifts is
strongly reduced, so that the actual trajectory of the flake center of mass
remains very close to the support scanline.
For example, springs of $K=2.5$~eV/\AA$^2$ yield perpendicular shifts $\leq
0.2$~\AA\ along the same trajectory.
The rotational freedom plus the lateral shifts of the flake can lead to
effectively commensurate contacts even for an incommensurate stacking, thus
explaining the deep energy valleys of the soft-spring pattern of
Fig.~\ref{AllVSxtip}b, eventually responsible for the stick-slip motion
demonstrated by the lateral force patterns of Fig.~\ref{AllVSxtip}a.
For hard springs constants, locking into deep energy minima does not occur,
but at the same time the flake is driven into highly repulsive geometries
which it cannot dribble.
This leads to higher force peaks, and eventually to a larger static
friction.

\subsection{Angular dependence of friction}

As a reference benchmark we consider a nearly rigid flake model, where
atoms are allowed to relax only in the $z$ direction, corresponding to the
$K\to\infty$ limit of the model studied until here, and comparing directly
to the model used by Verhoeven {\it et al.}\ \cite{Verhoeven04}.
The complete suppression of angular fluctuations should produce an
extremely sharp angular dependency of the friction force.

\begin{figure} 
\centering
\epsfig{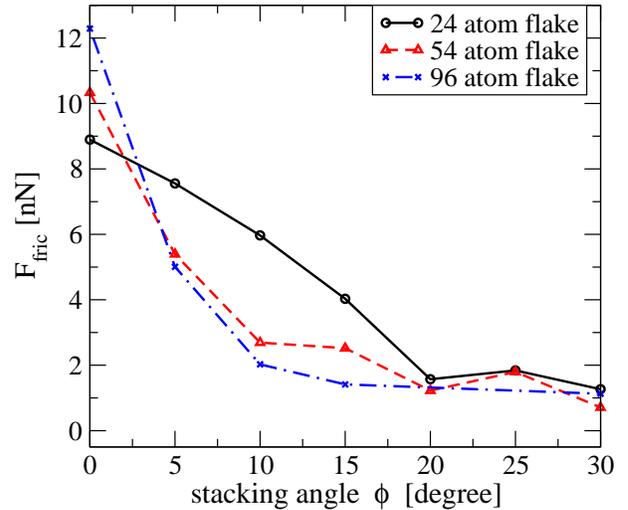}
\caption{(Color online) \label{FricVSphirigid}
Friction force $F_{\rm fric}$ as a function of the fixed stacking angle
$\phi$ for three different flake sizes: $24$ atoms (black solid line), $54$
atoms (red dashed line) and $96$ atoms (blue dot-dashed line). The
simulations are carried out with pulling angle $\theta=0^\circ$ and total
applied load of $100$~nN. Flake atoms are allowed to relax only along $z$
direction.
}
\end{figure}

Figure~\ref{FricVSphirigid} shows the computed static friction force as a
function of the fixed stacking angle $\phi$ for different flakes.
We note that friction decreases with the $\phi$ angle, showing a maximum
peak centered at $\phi=0^\circ$, similar to the outcome of previous model
calculation \cite{Verhoeven04}.
As shown by experiments \cite{Dienwiebel04,Dienwiebel05}, friction is
maximum at an highly commensurate contact ($\phi=0^\circ$) and decreases
rapidly as the flake rotates to incommensurate stackings.
The friction peak is sharper for wider flakes.
The sharpest peak for the 96-atom flake is similar to the one exhibited by
the rigid model of Ref.~\cite{Verhoeven04}.
$F_{\rm fric}$ decreases by nearly one order of magnitude from the
high-friction $\phi=0^\circ$ commensurate angle to the low-friction $\phi
\simeq 30^{\circ}$ incommensurate one.
This drop is smaller than was found by experiment \cite{Dienwiebel04}, where
it exceeded significantly one order of magnitude.
Also the absolute values of friction are systematically larger than experiment.
Experiment shows friction peak values near 0.2~nN, while the present model
yields a peak value of order 10~nN, 50 times larger.
This difference is even larger in the ``superlubric'' region near
$\phi=30^\circ$.
These differences are to be ascribed to the larger load, the adopted
short-ranged TB parameterization, and the neglect of thermal fluctuations,
as discussed below.

\begin{figure} 
\centering
\epsfig{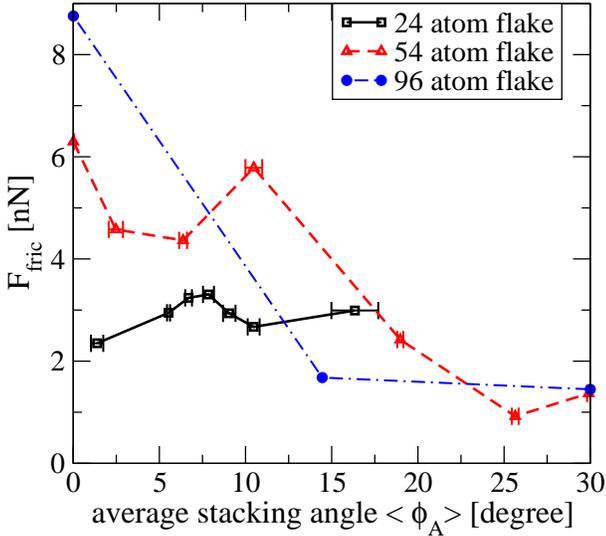}
\caption{(Color online) \label{FricVSang24-54}
The friction force $F_{\rm fric}$ as a function of the average rotation
angle $\langle \phi_A \rangle$ for three different flake sizes: $24$ atoms
(black solid line), $54$ atoms (red dashed line) and $96$ atoms (blue
points).
Simulations involve a pulling angle $\theta=0^\circ$, total applied load of
$100$~nN and soft springs constants $K=0.1$~eV/\AA$^2$.
}
\end{figure}

The rigid-flake models, studied here and in previous work
\cite{Dienwiebel04} do not look not especially realistic, since in practice
a carbon flake does deform, shift and rotate while interacting with the
graphite substrate and the AFM tip.
Figure~\ref{FricVSang24-54} reports the dependence of friction force
$F_{\rm fric}$ on the average rotation angle $\langle \phi_A \rangle$, for
a flake whose atoms are allowed to relax in all directions, for soft spring
constants $K=0.1$~eV/\AA$^2$.
In all calculations except those of 96-atom flake we use the same angles
$\phi=0^\circ$, $5^\circ$, $10^\circ$, $15^\circ$, $20^\circ$, $25^\circ$
and $30^\circ$, but the possibility of flake rotations allowed by the soft
springs produces significantly different effective average angles $\langle
\phi_A \rangle$, especially for the 24-atom flake.
Not surprisingly, with its vast rotational freedom, the 24-atom flake
displays an almost $\phi$-independent, constant friction.
For such a small flake with soft springs, rotations and shifts are so
effective to hinder the possibility to observe any reliable
$\phi$-dependency of $F_{\rm fric}$.
The 54-atom flake and, more clearly, the 96-atom flake show average angles
nearer to the support values, with smaller-amplitude fluctuations, and
therefore display a friction curve behavior with a peak at $\langle \phi_A
\rangle\simeq0^\circ$, fairly similar to the one obtained in the semi-rigid
case and observed in experiment, and with smaller friction at
incommensurate angles.
These results suggest that when the FFM tip happens to bind to a graphene
flake constituted by substantially less than approximately $10^2$ atoms, no
clear angular dependency and no superlubric regimes are observed.

\begin{figure} 
\centering
\epsfig{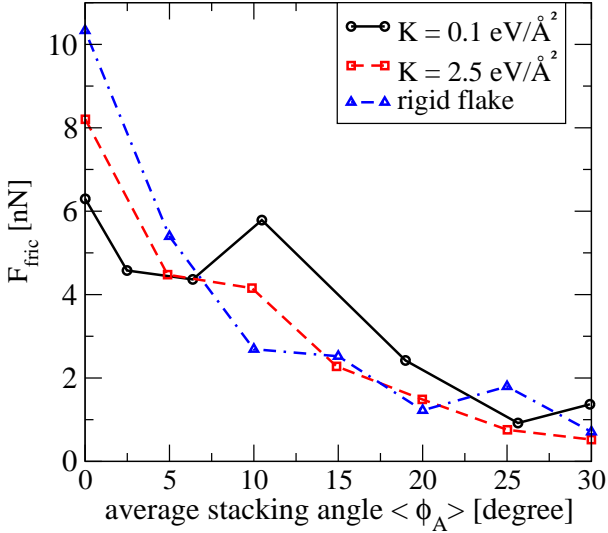}
\caption{(Color online) \label{FricVSang54k1-25}
Comparison of the friction force dependence on the average rotation angle
$\langle \phi_A \rangle$ for a 54-atom flake for two values of springs
strength: soft $K=0.1$~eV/\AA$^2$ (black solid line) and hard
$K=2.5$~eV/\AA$^2$ (red dashed line), plus the semi-rigid case (blue
dot-dashed line).
Simulations involve a pulling angle $\theta=0^\circ$, total applied load of
$100$~nN.
}
\end{figure}

Figure~\ref{FricVSang54k1-25} summarizes the effect of increasing the
rigidity of the springs on the friction dependency on $\langle \phi_A
\rangle$: the friction peak at a commensurate arrangement becomes sharper
and sharper as the spring rigidity increases.
At variance with the radical changes in $\langle \phi_A \rangle$ dependency
of the 24-atom flake as a function of the spring constant, for the 54-atom
flake the shift-rotational effects become comparably less important,
suggesting that for realistically large flakes in excess of one hundred
atoms, the precise value of the spring stiffness should become irrelevant,
as long as it remains in the $\lesssim 1$~eV/\AA$^2$ region.

\subsection{Load dependency}

\begin{figure} 
\centering
\epsfig{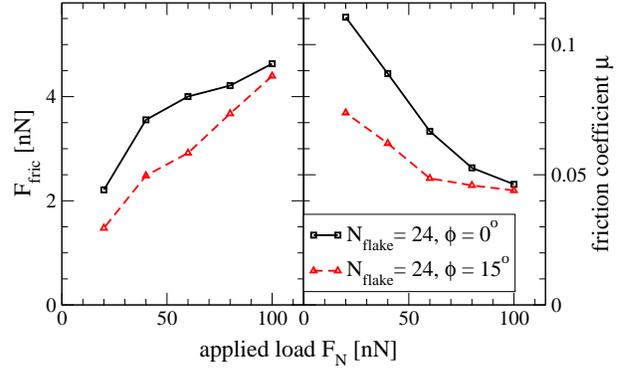}
\caption{(Color online) \label{FricVSload}
Nonlinear dependency of (a) the friction force and (b) the friction
coefficient $\mu=\frac{F_{\rm fric}}{F_N}$ on the total applied load $F_N$,
for a commensurate contact ($\phi=0^\circ$, black solid line) and an
incommensurate contact ($\phi=15^\circ$, red dashed line) of the 24-atom flake
flake.
Simulations are carried out for pulling angle $\theta=0^\circ$ and rigid
springs, $K=2.5$~eV/\AA$^2$.
}
\end{figure}

We come now to study the dependence of the friction force $F_{\rm fric}$ on
the applied load $F_N$, exploring a range 20 to 100~nN, matching typical
experiment values \cite{Dienwiebel04,Sasaki02}.
Figure~\ref{FricVSload} shows the dependence of the friction force $F_{\rm
  fric}$ and coefficient $\mu\equiv F_{\rm fric} / F_N$ on the applied
load.
Hard springs are selected to reduce the flake shift-rotational effects, in
order to focus on the load dependence of friction and simpler comparison
with earlier results.
Friction increases with load, but significant deviations from the linear
Coulomb law are observed, especially for commensurate stacking
$\phi=0^\circ$.
Observe that experiment found an even weaker dependency of the friction
force on load \cite{Dienwiebel04}.
Although the data do not point clearly in the direction of a power-law
behavior $F_{\rm fric}\propto F_N^\alpha$, it is clear that if any such law
was to be estimated, it would have $\alpha<1$.
This is at variance with previous findings for a sharp undeformable
tip-surface contact \cite{Fusco04}, and with recent studies of the sliding
of hydrogen-passivated carbon \cite{Mo09}.
The resulting static friction coefficient approaches the standard
macroscopic value \cite{CRC94} of graphite-graphite contact, $\mu=0.1$,
while much smaller values are found in the single-crystal FFM experiments
addressed by the present model.
Regardless of the applied load, the flake-substrate distance being smaller
in the model than in real life produces larger absolute values of friction,
and this overestimation is particularly severe at small load.

\subsection{Scanline dependency}

\begin{figure} 
\centering
\epsfig{file=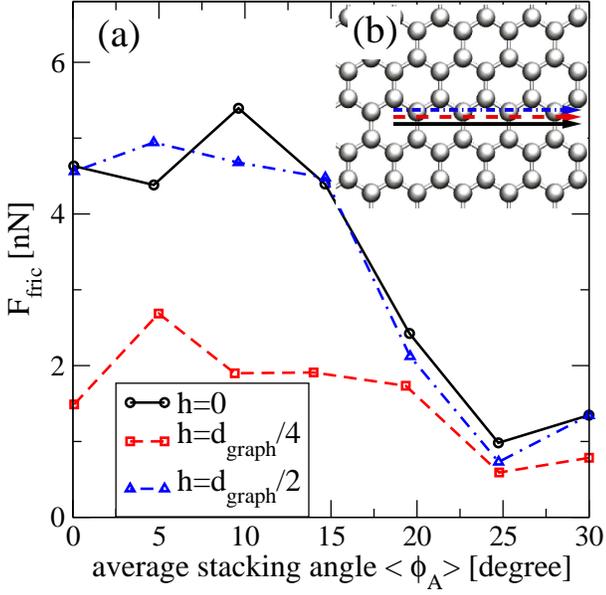,width=80mm,clip=}
\caption{(Color online) \label{FricVSang-scanline}
(a) Friction force $F_{\rm fric}$ as a function of the average stacking
  angle $\langle \phi_A \rangle$ for three different scanlines drawn in
  panel (b), defined by the three following initial stackings of the
  support over the substrate: AB (black solid), AB with a transverse shift
  $h=d_{\rm graph}/ 4$ (red dashed) and AB with transverse shift
  $h=d_{\rm graph}/ 2$ (blue dot-dashed).
The simulations involve a 24-atom flake, pulling angle $\theta=0^\circ$,
applied load $F_N=100$~nN and hard springs of constants $K=2.5$~eV/\AA$^2$.
}
\end{figure} 

We investigate the dependence of the friction force versus stacking angle
on the actual scanline followed by the support.
Changing scanline determines a different potential profile seen by the
flake, thus modifying the frictional behavior \cite{Fusco04}.
Figure~\ref{FricVSang-scanline} reports the friction angular dependency for
three equally spaced scanlines.
We carry out simulations for hard springs $K=2.5$~eV/\AA$^2$ where scanline
effects are the most visible, because of hindered lateral shifts.
For the $h=0$ and $h=d_{\rm graph} / 2$ scanlines we find a similar
friction for all values of the stacking angle $\langle \phi_A \rangle$,
while the $h=d_{\rm graph}/ 4$ scanline shows systematically lower
friction, especially for small $\phi$.
The reason is that for $\phi_A\simeq 0$ along this special line each flake
atom never hits any substrate atom directly on top, therefore effectively
finding a significantly lower corrugation.
In contrast along the two other scanlines, for $\phi=0^{\circ}$ one half of
flake atoms encounters periodically a substrate atom right below its
trajectory, thus finding a high corrugation.
Softer springs produce a much weaker dependence on the scanline: the flake
takes advantage of its freedom to displace laterally, thus following
low-corrugation lines (such as $h=d_{\rm graph}/4$) even when the
support pulls it along some nearby parallel line.

\begin{figure} 
\centering
\epsfig{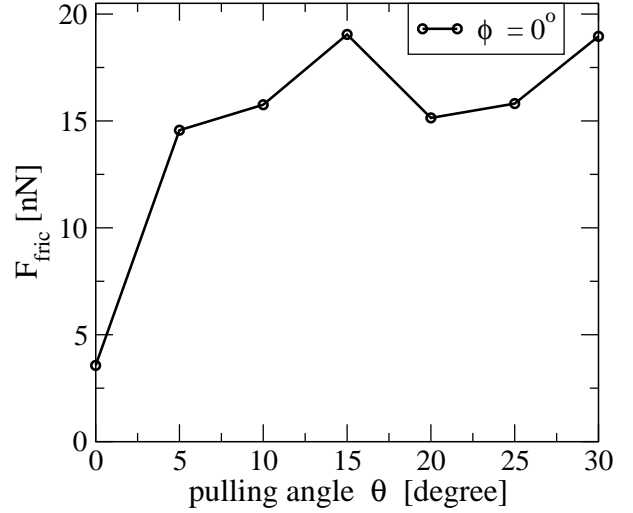}
\caption{\label{FricVStheta}
Friction force $F_{\rm fric}$ as a function of the pulling angle $\theta$.
Static friction data are obtained by averaging three different scanlines,
defined by initial stacking AB, AB shifted by $d_{\rm graph} / 4$ and AB shifted by
$d_{\rm graph} / 2$ perpendicular to the pulling direction.
Simulations involve a 24-atom flake with load of $100$~nN, support
stacking angle $\phi=0^\circ$ and springs constants $K=2.5$~eV/\AA$^2$.
}
\end{figure}

The scanlines of Fig.~\ref{FricVSang-scanline} involves $\theta=0^\circ$,
i.e.\ a pulling along the $x$ direction, where the flake encounters
periodic repetitions of the substrate potential.
A different pulling angle affects directly this periodicity of the problem,
in general leading to a nonperiodic profile.
Figure~\ref{FricVStheta} displays the friction force as a function of the
pulling angle $\theta$.
Data are averaged over three different scanlines, defined by initial
stacking with $h=0$, $h=d_{\rm graph} / 4$ and $h=d_{\rm graph} / 2$.
Like in previous calculations \cite{Verhoeven04}, we find a minimum
friction for pulling angle $\theta=0^\circ$, followed by a fast growth in
friction (until $\theta=10^\circ$).
We attribute the observed differences between our results and those by
Verhoeven {\it et al.}\ to the different interaction models.

\section{\label{conclusion} Discussion and conclusion}

We find fair qualitative agreement between the results  obtaining by our
TB atomistic model and the existing experimental data, with a few
significant differences.
Firstly, our calculations recover the stick-slip behavior of the lateral
forces, characteristic of FFM sliding experiments.
In particular, we find the correct qualitative dependence of stick-slip on
the springs stiffness characterizing the cantilever-tip-flake coupling:
soft springs allow for clean stick-slip behavior, while hard springs
inhibit it.
Our calculations also reproduce correctly the friction pattern as a
function of the average stacking angle $\langle \phi_A \rangle$ especially
as long as the rotational degree of freedom $\phi_A$ is quenched.
We also find that for larger flakes, the fluctuation in $\phi_A$ are
suppressed automatically anyway, due to the larger torque exerted by the
springs connecting the flake to the tip.
Accordingly, for flakes of sufficiently large size incommensurability
produces significantly less friction, although the friction drop is
smaller than in experiment.
In the quantitative comparison between the experimental results and our
model, we find static friction force $F_{\rm fric}$ and coefficient $\mu$
systematically at least one order of magnitude larger than experiment, this
difference being especially significant in the incommensurate
configurations where no proper superlubric regime is observed.
These and other quantitative discrepancies are to be attributed to:
(i) The reduced interlayer equilibrium distance, related to the small
cut-off distance of the present TB parameterization, which is responsible
for the increased energy corrugation experienced by our model flake with
respect to real graphene on graphite.
(ii) The extra reactivity of the isolated model graphene flake with respect
to a real one, which is bond to the AFM tip and thus somewhat passivated;
accordingly, especially the atoms at the flake border, show a greater
tendency to react with substrate atoms, thus increasing friction.
(iii) The neglect of thermally-activated slips through energy barriers
\cite{Riedo03,Gnecco03}: this neglect generates an overestimation of
friction especially where these barriers are lower, i.e.\ at incommensurate
stackings.
Indeed the current understanding \cite{Frenken09conf} of the observed
\cite{Dienwiebel04} superlubric sliding involves thermolubricity associated
to a high attempt rate for overcoming the corrugation barriers due to the
microscopic mass of the vibrating tip apex.
If the experiment of Ref.~\cite{Dienwiebel04} could be repeated at the much
lower temperature of a few degree Kelvin, the observed friction values and
dependency on the $\Phi$ angle would probably look much more similar to
the one obtained in the present model.

Calculations carried out with comparably soft springs and small flakes
($N_{\rm fl}\leq 54$) show that the flake shift-rotational freedom
increases friction for incommensurate stackings (by allowing the flake to
explore deeper-bound minima) and decreases it for commensurate ones (by
allowing the flake to dribble the top barriers): the result is a
substantial flattening of the dependency of the friction static force
$F_{\rm fric}$ on the stacking angle $\phi$.
Harder springs (e.g.\ $K=2.5$~eV/\AA$^2$) would suppress the flake freedom
to rotate and shift laterally but are incompatible with the clear
stick-slip behavior observed experimentally.
These considerations confirm that the size of the flakes showing
superlubric sliding in actual FFM experiments is large $N_{\rm fl}\geq 96$.
Many discrepancies with experiment would probably be disposed of, if a
longer ranged interatomic interaction was employed, for example a TB model
based on a longer cutoff \cite{Papaconstantopoulos98}.
This way, a much weaker flake-surface interaction would effectively
correspond to comparably stronger tip-flake interaction, thus a significant
hindering of rotations and translations even with a realistically weak
tip-flake coupling $K\leq0.5$~eV/\AA$^2$.
If accurate long-range C-C interactions up to distances of the order of
1~nm were present in the force field, one could even include substrate
relaxation to study an even more realistic model.
Such an improved model would however imply significantly larger
computational workload, especially if thermal excitations were also
included.
Further research should also investigate the effect of the presence of
structural defects in the flake or in the substrate, as proposed by Guo
{\it et al.}\ \cite{Guo07}, and the effect of flake shape.

\section*{Acknowledgments}
We are grateful to J.\ Frenken and R.\ Buzio for useful discussion.
We acknowledge financial support by the project MIUR-PON ''CyberSar'',
and by the Italian National Research Council (CNR) through contract
ESF/EUROCORES/FANAS/AFRI.

\end{document}